\documentclass[aip,reprint]{revtex4-1}

\usepackage{amsmath,amssymb}
\usepackage{graphicx}
\usepackage{dcolumn}
\usepackage{bm}

\usepackage[utf8]{inputenc}
\usepackage[T1]{fontenc}
\usepackage{mathptmx}
\usepackage{etoolbox}
\usepackage{color}

\begin{document}

\title{Pervaporation-driven electrokinetic energy harvesting\\ using  poly(dimethylsiloxane) microfluidic chips}

\author{Hrishikesh Pingulkar}
\affiliation{CNRS, Syensqo, LOF, UMR 5258, Université de Bordeaux, 178 av. Schweitzer,  33600 Pessac, France.}%

\author{Cédric Ayela}%
\affiliation{Université de Bordeaux, IMS, CNRS, Bordeaux INP, UMR 5218, 33607 Pessac, France.}%

\author{Jean-Baptiste Salmon}
\affiliation{CNRS, Syensqo, LOF, UMR 5258, Université de Bordeaux, 178 av. Schweitzer,  33600 Pessac, France.}%
\email{Jean-Baptiste.Salmon-exterieur@syensqo.com}


\begin{abstract}
Electrokinetic energy harvesting from  evaporation-driven flows in porous materials has recently been the subject of numerous  studies, particularly with the development of nanomaterials with high conversion efficiencies. The configuration in which the energy conversion element is located upstream of the element which passively drives the evaporative flow has rarely been studied. However, this configuration offers the possibility of increasing the harvested energy  simply by increasing the evaporation surface area and/or the hydraulic resistance of the energy conversion element. In this work, we investigate this configuration with poly(dimethylsiloxane) (PDMS) chips playing the role of  {\it artificial leaves} driving a pervaporation-induced flow through a polystyrene colloid plug in a submillimetre tube for  the energy conversion. With an appropriate design of the venation of the PDMS leaves, we report the first experimental evidence of electrokinetic energy conversion from pervaporation-induced flows, which increases with the pervaporation area. We also provide new insights by demonstrating that this increase is limited by cavitation within the PDMS leaves, which occurs systematically as soon as the water pressure inside the leaf reaches $P_\text{leaf} \simeq 0$~bar. Whatever the cavitation threshold, this phenomenon imposes an intrinsic limit on this configuration, underlining the need for innovative strategies to improve the harvesting of  electrokinetic energy by evaporation.
\end{abstract}

\maketitle


\section{Introduction}\label{sec:intro}
Since Osterle's pioneering work~\cite{Osterle1964}, electrokinetic (EK) mechanisms for converting mechanical energy into electrical energy, based on the concepts of streaming currents and streaming potentials in pressure-driven flows, have been well established both experimentally and theoretically~\cite{Olthuis2005,Lu2006,VanDerHeyden2007,VanDerHeyden2006,Mansouri2012,Zhang2020}. 
Most of these studies focus on the EK energy conversion efficiency, because  the  mechanical input energy associated with the flow is imposed externally in the  intended applications (hydrostatic pressure).

More recently, it has been shown that water evaporation from saturated porous media can also be used to harvest electrical energy using the same EK mechanisms~\cite{Li2017,Xue2017}. These EK mechanisms also partly explain the harvesting of electrical energy from ambient humidity, as recently demonstrated using various specific nanoporous materials, commonly known as {\it moist-electric generators}~\cite{Liu2020,Liu2022,Shen2020}.  Evaporation-induced flows occur spontaneously and continuously because they are passively driven by chemical potential gradients, thus opening up new application possibilities. These studies are part of a more general theme that considers the natural evaporation of water, as well as the ambient moisture, as  potentially reliable sources of renewable energy~\cite{Shen2020,Cavusoglu2017,Zhang2018}. To date, several experimental demonstrations of EK energy conversion from water evaporation in saturated porous media  have been reported, involving energy conversion materials as diverse as cellulose-based paper~\cite{Das2018} and  textiles~\cite{Das2019}, nanoporous membranes~\cite{Saha2021}, glass microchannels~\cite{Liu2023,Yanagisawa2023}, clay minerals~\cite{Bora2024}, carbonized biomass~\cite{Jiao2023},
but also electronically conducting and carbon-based nanomaterials~\cite{Ding2017,Zhang2019,Yun2019,Bae2020,Qin2020,Bae2022} for which the energy conversion mechanisms involve more subtle effects~\cite{Liu2018}.  
Most of these studies are mainly proofs of concept focusing on EK energy conversion performance via the development of new materials. 

\begin{figure}[htbp]
\centering
\includegraphics{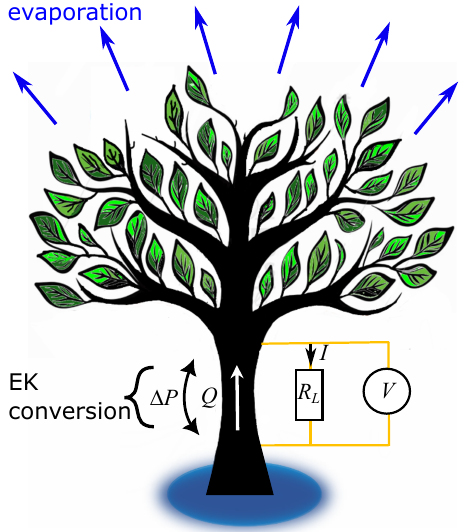}
\caption{Illustration of evaporation-driven EK energy harvesting in a configuration for which the evaporation surface (\textit{leaves}) and the EK conversion element (\textit{trunk}) are separate and in series.
By analogy with evapo-transpiration in plants, evaporation  induces a flow at a rate $Q$,  associated to a pressure drop $\Delta P$ across the EK conversion element. A fraction of the streaming current is collected in a load resistance $R_L$ placed in parallel. 
}\label{fig:tree}
\end{figure}
Yaroshchuk recently carried out a critical review of this rapidly expanding field~\cite{Yaroshchuk2022}, and pointed out that, to date, there has been no satisfactory quantitative comparison with theoretical models, even in the case of standard EK mechanisms, notably due to the lack of precise experimental characterisation of the electrochemical collection of the streaming currents by the electrodes. Yaroshchuk also identified that the configuration in which the element driving the flow  is separate from the EK conversion offers better optimisation possibilities. This configuration is shown in Fig.~\ref{fig:tree} thanks to the analogy with evapo-transpiration that continuously transports water in plants, from the soil to the leaves to maintain their hydration~\cite{Stroock2014,Jensen2016}. 
In this illustration, evaporation from the leaves drives a flow at a rate $Q$ through 
an EK conversion element (\textit{trunk})  which  harvests
 electrical power $\mathcal{P}_e = V I$ from the input mechanical power $\mathcal{P}_h= \Delta P Q$  with an efficiency $\epsilon = \mathcal{P}_e/\mathcal{P}_h$,
$I$ being the current collected in a parallel load resistance $R_L$, $V$ the associated electric potential, and $\Delta P$ the pressure drop across the EK conversion element. 
As pointed out by Yaroshchuk~\cite{Yaroshchuk2022}, it is the flow rate $Q$ and the hydraulic resistance $R_h = \Delta P/Q$ of the EK conversion element that are the relevant parameters in such  passive configurations for significantly increasing the electrical power, as  $\mathcal{P}_e = \epsilon R_h Q^2$. 
To our knowledge, this configuration has only been explored by a few groups with the emphasis on the possible applications~\cite{Li2017,Das2019,Liu2023}.

\begin{figure}[ht!]
\centering
\includegraphics{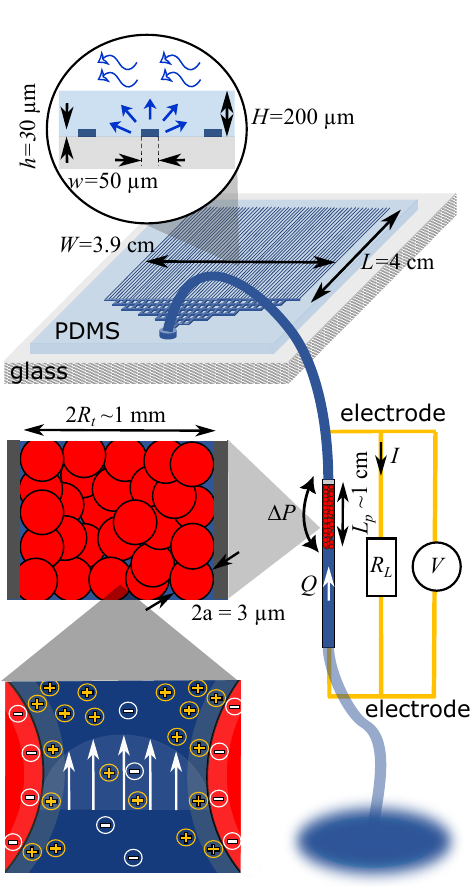}
\caption{EK energy harvesting from  pervaporation-driven flows in a PDMS leaf. The EK conversion element consists of a plug of charged-stabilised colloids upstream of a filter fitted into a tube. Pervaporation in the PDMS leaf induces a flow rate $Q$ and streaming currents through the interstices of the plug that are  collected by electrodes in a load resistance $R_L$. The output electric power is $\mathcal{P}_e = V I$ and the input mechanical power is $\mathcal{P}_h = \Delta P Q$. 
}\label{fig:PervapPrinciple}
\end{figure}

The aim of our work is firstly to provide a proof of concept of this configuration using {\it artificial  leaves} made in poly(dimethylsiloxane) (PDMS), see Fig.~\ref{fig:PervapPrinciple}, and then to study its intrinsic limitation due to cavitation, similar to the embolisms observed in trees during periods of intense drought~\cite{Stroock2014}.  PDMS not only enables the rapid prototyping of  chips using soft lithography, but is also slightly permeable to water. Water molecules from microchannels in a PDMS chip thus solubilise in the matrix,
diffuse, and evaporate into ambient air, a process known as {\it pervaporation}. Due to mass conservation, pervaporation inevitably induces flows in
PDMS chips,  at rates $Q$ of the order of a few nL\,h$^{-1}$ for a single  channel of a few centimetres long~\cite{Verneuil:04,Randall:05}.    
    Although these flows are often negligible for the majority of microfluidic studies, they have also found numerous applications~\cite{Bacchin2022}.
In the present work, we optimized the cm$^2$  channel network of the  PDMS leaves~\cite{Noblin:08} to achieve  pervaporation-driven flow rates up to $Q \simeq 20~\mu$L\,h$^{-1}$.
For the EK conversion element, we followed the strategy proposed in Refs.~\cite{Saha2019,Chen2009} and  fabricated a plug of charged-stabilized colloids closely-packed  upstream of a filter in a submillimetre-sized tube. Pressure-driven flow through this charged nanoporous medium then results in a streaming current that can be collected in an external electrical circuit. We finally tuned  the hydraulic resistance $R_h$ of the colloid plug  to reach relatively large pressure drops $\Delta P = R_h Q$ from the pervaporation-driven flow rates, and hence a 
measurable output electrical power $\mathcal{P}_e$ that varies with the area of the PDMS leaves. We nevertheless  showed that this output energy is limited by cavitation in the PDMS leaves, as soon as the 
pressure drop reaches $\Delta P \simeq 1$~bar. Whatever the cavitation threshold that would be reached for any other microfluidic system, our results show that cavitation in the configuration shown in Fig.~\ref{fig:tree} possibly 
limits the strategy of increasing the evaporation surface to increase the energy harvested.

\section{Materials and methods\label{sec:MaterialsMethods}}
\subsection{Fabrication of PDMS leaves\label{subsec:fabPDMSleaves}}

Firstly, we used standard photolithography techniques  to obtain a mold of microchannels with height $h=30~\mu$m on a silicon wafer  (SU-8 photoresist).
The design of the network of the channels  is shown in the 3D view in Fig.~\ref{fig:PervapPrinciple} (see also Fig.~S1 in Supplementary Material). It consists of $N=79$ parallel dead-end channels of length $L=4~$cm connected to a single inlet. The width of each channel is $w=50~\mu$m, the center-to-center distance between adjacent channels is $d = 500~\mu$m, leading to a width of $W=3.9$~cm for  the entire network.

We then used soft lithography techniques to make PDMS artificial leaves  from this network of channels. 
More precisely, a PDMS layer of thickness $H \simeq  200~\mu$m (Sylgard 184, curing agent/PDMS ratio: $1/10$) is initially spin-coated onto the mold and then cross-linked at $65^\circ$C for a few hours. At the same time, a thicker PDMS structure ($\simeq  5$~mm) with a hollow square cavity slightly larger than the  area $L\times W$ of the channel network is made on a bare silicon wafer.  This PDMS block with a frame-like structure is then peeled off from the wafer and bonded to  the thin PDMS layer covering the channels using a plasma treatment.
The assembly is then  carefully peeled off from the wafer, punched to make the fluid inlet, and bonded to a glass slide using again a plasma treatment. The thick PDMS frame is essential not only to ensure the mechanical strength of the tube connected to the fluid inlet of the channels, but also to 
 handle the thin cm$^2$ PDMS layer during the peeling step and avoid wrinkles that would trap air bubbles during the bonding step.

\subsection{Electrical and flow measurements \label{subsec:ElectricalFlowMeasurements}}
Fluid pressure was controlled in the experiments described below using the Fluigent MCFS-EZ device, capable of imposing pressure differences of up to $\Delta P = 7$~bar. 
Flow rates were measured using the flow units Fluigent S ($0 \pm 7~\mu$L\,min$^{-1}$) and XS ($0 \pm 1.5~\mu$L\,min$^{-1}$) depending on the flow rate range.
Streaming potentials $V$ were recorded using a digital multi-meter (Agilent, 34405A) with an internal impedance of $R_\text{app}= 10$~M$\Omega$.
The resistance $R_\text{var}$ in parallel of the EK conversion was varied in the  $R_\text{var} = 1~\Omega$ to 10~M$\Omega$ range using a decades resistance box (Centrad, DR07). In this range, the impedance of the multi-meter must be taken into account and the load resistance $R_L$ in Fig.~\ref{fig:PervapPrinciple} is  $R_L = R_\text{var} R_\text{app}/(R_\text{var}+ R_\text{app})$. Local observations of the PDMS leaf were obtained using a stereo-microscope (Olympus SZX10) coupled with a sCMOS camera (Hamamatsu).

\subsection{EK conversion element\label{subsec:EKconversionelement}}

\begin{figure}[htbp]
\centering
\includegraphics{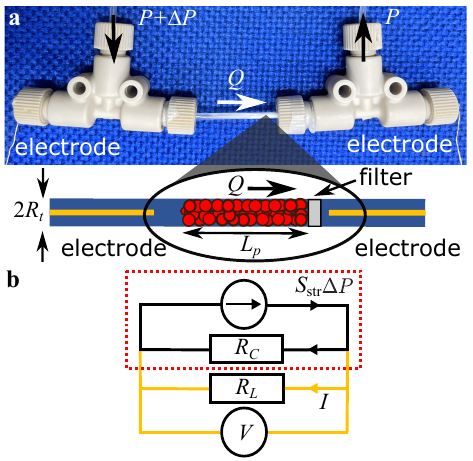}
\caption{EK conversion element. (a) Millifluidic assembly  showing the  colloidal plug  upstream of a filter fitted into a tube of inner diameter $2R_t$  ranging from $0.25$ to $1$~mm depending on the experiments. Ag/AgCl electrodes located close to the plug and downstream of the filter are used to collect the  current. 
$\Delta P$ is the  pressure drop 
 across the plug. 
 (b) Equivalent circuit of the EK conversion element (red dotted rectangle) connected to a load resistance $R_L$ (see text).
    }\label{fig:EKconversionElement}
\end{figure}

We used polystyrene latex beads of mean diameter $2a = 3~\mu$m 
(Sigma Aldrich, LB30, batch mass fraction $10$$\%$) as the building block of the colloidal plug of the EK conversion element shown in Fig.~\ref{fig:EKconversionElement}(a). We first prepared dispersions of known volume fraction~$\varphi_0$ by dilution of the batch dispersion with deionised water,
 typically $\varphi_0=5$\%.
The particles are then accumulated by frontal filtration 
 in a submillimetre-sized tube (inner diameter $2R_t$ ranging from $0.25$ to $1$~mm depending on the experiments, IDEX) upstream of a filter (porosity $0.5~\mu$m, IDEX, P-273x) nested in a fluidic T connector (PEEK, 
 IDEX), see Fig.~\ref{fig:EKconversionElement}(a). More precisely, a given volume $V_c$ of dispersion is injected into the tube, then filtered at a typical pressure of $6$~bar with water upstream, while measuring simultaneously  the flow rate. We consider that the colloidal plug is correctly formed as soon as the flow rate reaches a steady value.
 The desired length $L_p$ of the plug is estimated from  mass conservation $\varphi_0 V_c \simeq \pi R_t^2 L_p \varphi_d$ assuming the particles are randomly close-packed ($\varphi_d \simeq 0.64$), and also measured directly in the tube using a ruler.

To collect streaming currents, we used 
Ag/AgCl electrodes prepared from silver wires of diameter $500~\mu$m (Sigma-Aldrich). 
Wires of $10$~cm long are first rinsed with isopropanol to remove any surface contamination, then partially immersed in a $30$\% NaClO (Clorox bleach) solution for $30$~min to chloride the silver wires.
The prepared electrodes are then rinsed with isopropanol and deionised water and are  embedded within the T connectors, as shown in Fig.~\ref{fig:EKconversionElement}(a). 
This configuration makes it easy to adjust the position of the electrodes, one upstream and close to the colloidal plug, the other downstream and close to the filter.

In our experiments, the colloidal particles are negatively charged so that the flow advects predominantly cations. Chloride ions are thus released at the right electrode in Fig.~\ref{fig:EKconversionElement} (cathode) due to the transformation of AgCl into metallic Ag (AgCl(s) $+$
e$^-$ $\to$ Ag(s) $+$ Cl$^{-}$), while chloride ions are integrated in the left electrode in Fig.~\ref{fig:EKconversionElement}  (anode), therefore leading to a flux of electrons in the external electrical circuit from the anode to the cathode, see Ref.~\cite{Yaroshchuk2022} for more information. 

\section{Results and discussion}\label{sec:Resultsdiscussion}

\subsection{Pervaporation-driven flows: from a single channel to parallel leaves \label{sec:secpervap}}
    We first discuss the design and characterisation of the PDMS leaf illustrated schematically in Fig.~\ref{fig:PervapPrinciple}, which passively drives a flow due to water evaporation in ambient air. 
 Dollet {\it et al.}\ showed that the pervaporation-driven flow rate $Q_i$ for a  single dead-end channel of length $L \gg H \gg h$ is given by~\cite{Dollet2019}:
\begin{eqnarray}
Q_i =  \tilde{q} F L  (1-\text{RH}), 
\label{eq:1DchannelTheory}
\end{eqnarray}
where  RH is the ambient relative humidity and 
$F$ a geometrical factor. 
$\tilde{q}$ is related to the solubility and diffusivity of water in PDMS and of the order of $\tilde{q} \simeq 0.5~\mu$m$^2$\,s$^{-1}$~\cite{Verneuil:04,Randall:05,Dollet2019,Harley2012,Bacchin2022}.
Since $\tilde{q}$ is an intrinsic parameter of PDMS, the only possible optimisation in our application  is geometric ($F$ and $L$). 

\begin{figure}[htbp]
\centering
\includegraphics{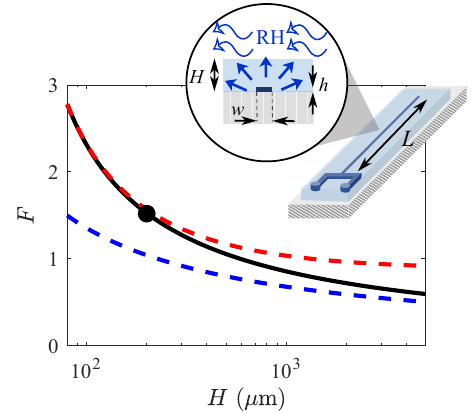}
\caption{Schematic perspective and sectional views  of a single channel in a PDMS chip.
Geometrical factor $F$ in eqn~\ref{eq:1DchannelTheory} for the case $h=30~\mu$m and $w=50~\mu$m calculated   by Dollet {\it et al.}~\cite{Dollet2019}. The red dotted line is  eqn~S2 which correctly approximates the analytical solution for $H\leq 200~\mu$m. The blue dotted line is eqn~S1 which roughly approximates the case of a thick chip. The black dot indicates the value $F \simeq 1.5$ for $H=200~\mu$m.
}\label{fig:SingleTheoreticalPervap}
\end{figure}

 Dollet {\it et al.}\  also provided exact analytical solutions for $F$ using conformal-mapping techniques for  a channel with 
rectangular cross-section~\cite{Dollet2019}.
Fig.~\ref{fig:SingleTheoreticalPervap} shows this calculation (as well as approximations detailed in Supplementary Information, Sec.~S2) for the case $h=30$,  $w=50~\mu$m and various thicknesses $H$. These calculations show that the geometrical factor  increases only moderately  from   $F  \simeq 0.6$ for $H=5$~mm to $F\simeq 2.3$ for a $H=100~\mu$m thin chip. 
Higher factors could be achieved for thinner chips (because of the divergent term in eqn~S2 when $H \to h$), but the handling  of thin cm$^2$ PDMS layers in the microfabrication stages remains tricky, and we found that $H\simeq 200~\mu$m is a good compromise ($F \simeq 1.5$). Such thicknesses lead to
pervaporation-driven flow rates of the order of $Q_i \simeq 0.1~\mu$L\,h$^{-1}$ assuming 
 $\tilde{q} \simeq 0.5~\mu$m$^2$\,s$^{-1}$ and a $L= 4$~cm centimetre long channel (at $\text{RH}=0$).  As we will see later, such flow rates are too low for the proof of concept of EK energy conversion targeted by our study.

\begin{figure}[ht!]
\centering
\includegraphics{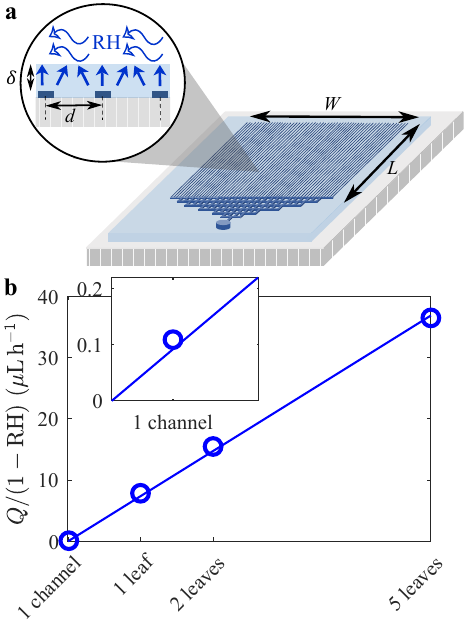}
\caption{(a) Schematic perspective and sectional views  of the PDMS leaf. The channels have a rectangular cross-section $h \times w$ and $\delta = H-h$, $H$ being the thickness of the PDMS layer. 
(b) Measured flow rate $Q$ achieved  using  multiple leaves placed in parallel. $Q$ is re-scaled by $1-\text{RH}$ ($\text{RH}$ varied from $0.4$ to $0.5$ in these experiments). The $1$-channel case corresponds to the theoretical estimation of eqn~\ref{eq:1DchannelTheory} for $H=200~\mu$m, and is also shown in the inset. The continuous line is a linear fit.}\label{fig:PervaporationLeaves}
\end{figure}

To reach much higher flow rates, the dead-end channels need to be parallelized in a single chip, as is the case for the venation of real leaves, see Fig.~\ref{fig:PervaporationLeaves}(a) showing $N$ parallel channels  connected to the same inlet.
This configuration has been studied by Noblin {\it et al.}~\cite{Noblin:08} who   showed that the  overall pervaporation rate  follows $Q = N Q_i$ for small channel density $1/d \ll 1/H$, but saturates for higher density
 to  the limiting  value:
\begin{eqnarray}
Q_\mathrm{lim} \simeq \frac{LW}{\delta} \tilde{q} (1-\text{RH})\,.\label{eq:pervap1D}
\end{eqnarray}
This regime is due to the 
screening of the diffusion of  water  in the PDMS matrix between the channels and corresponds to the  1D pervaporation through a PDMS layer of thickness $\delta=H-h$. 
This effect is similar to the screening of evaporation from a network of pores,  as in the case of the stomata of real leaves~\cite{Lehmann2015,Jensen2016}, or as for nanoporous media in the context of EK energy harvesting~\cite{Yaroshchuk2021}. The experimental case studied in this  work, $N=79$ parallel channels in a $W=3.9$~cm wide leaf,  lies a priori  between these two regimes because $d = 500~\mu$m is close to $H = 200~\mu$m. We thus used numerical resolutions detailed in Supplementary Information, see Fig.~S2 and S3, to show that the pervaporation-driven flow rate in our case is given by $Q = \alpha  Q_\mathrm{lim}$ with $\alpha \simeq 0.5$. 

Fig.~\ref{fig:PervaporationLeaves}(b) reports the measurements of the flow rate $Q$ induced by  pervaporation of water from $1$, $2$, and $5$~PDMS leaves connected in parallel.
In these experiments, the leaves are gently filled with pure water by imposing an inlet pressure of between $1$ to $3$~bar, to completely remove the air initially present in the dead-end channels. The flow rate $Q$ is then measured thanks to a flowmeter upstream of the leaves, and rescaled by $1-\text{RH}$ to account for the  variations of the ambient conditions (temperature range $\simeq 20$--22$^\circ$C, $\text{RH}=0.4$--$0.5$). Our data show a linear scaling of $Q/(1-\text{RH})$ with the pervaporation area as expected from theory.
The linear fit in Fig.~\ref{fig:PervaporationLeaves}(b) together with
eqn~\ref{eq:pervap1D} and the prefactor $\alpha \simeq 0.5$  estimated numerically leads to $\tilde{q} \simeq 0.4~\mu$m$^2$\,s$^{-1}$ in agreement with the reported values so far~\cite{Verneuil:04,Randall:05,Dollet2019,Harley2012,Bacchin2022}.
The maximal rate measured of the order of $Q\simeq 18~\mu$L\,h$^{-1}$ for the $5$-leaves case at $\text{RH}=0.5$, corresponds a $\simeq 350$-fold increase of the single channel case. 
As demonstrated later in this work, such values are now compatible with the proof of concept of EK conversion from pervaporation-driven flows.

The 1D description of eqn~\ref{eq:pervap1D} leads to the pervaporation flux per unit surface $J_p = \alpha \tilde{q}(1-\text{RH})/\delta$, which can be compared to the water evaporation flux  $J_e$ from a free surface, or equivalently from a dense array of  nanopores~\cite{Yaroshchuk2021,Yaroshchuk2022}. The later is given by $\rho_w J_e \simeq D_w^\text{air}C^\text{air}_\text{sat} (1-\text{RH})/l$, $\rho_w$ (kg\,m$^{-3}$) being the density of liquid water,
$C^\text{air}_\text{sat}$ (kg\,m$^{-3}$) the water concentration in the gas phase at saturation, $D_w^\text{air}$ (m$^2$\,s$^{-1}$)  the diffusion coefficient of water as vapor in the gas phase, and $l$ the thickness of the stagnant boundary layer. 
For $l \simeq 2~$mm, typical for moderate air convection over a surface area of a few square centimetres~\cite{Yaroshchuk2022}, $J_e \simeq 200$~nm\,s$^{-1}$ for ambient conditions ($\text{RH}=0.5$) while $J_p \simeq 1$~nm\,s$^{-1}$. 
$J_p \ll J_e$ is in line with the low water permeability of PDMS, and also demonstrates that the measured pervaporation rates shown in Fig.~\ref{fig:PervaporationLeaves}(b) 
do not depend on the air convection around the leaves, as the overall mass transfer is limited by the pervaporation through the PDMS matrix (resistance $\mathcal{R}_\text{PDMS} = \delta/(\alpha \tilde{q})$) and not by mass transfer in air (resistance
$\mathcal{R}_\text{air} = \rho_w l/(D_w^\text{air}C^\text{air}_\text{sat}) \ll \mathcal{R}_\text{PDMS}$).

For  a single leaf, the flow rate $Q \simeq 3.9~\mu$L\,h$^{-1}$ for $\text{RH}=0.5$ (see Fig.~\ref{fig:PervaporationLeaves}(b)) 
is associated with an emptying time of the channel network of the order of $\tau_p \simeq N w L h/Q \simeq 70$~min. Pervaporation  therefore induces water flows with an average velocity $V_p \simeq L/\tau_p \simeq 9~\mu$m\,s$^{-1}$ at the entrance of each channel, decreasing linearly  to $0$ at their end according to mass conservation~\cite{Dollet2019}.
This flow inevitably leads to a continuous accumulation in the microfluidic channels of all the solutes that are insoluble in the PDMS matrix  (e.g., salts), as also observed in the context of evaporation-driven electrokinetic energy conversion in microfluidic glass channels~\cite{Yanagisawa2023}, see also
Ref.~\cite{Bacchin2022}.
Because of the 
balance between pervaporation-driven advection and solute diffusion, the concentration of solutes only increases at the tip of the channels in a volume $\simeq p w h$, with $p \simeq  \sqrt{D_s \tau_p}$ ($D_s$ being the solute diffusion coefficient) and at a rate  
$p\,\dot{C}  \simeq V_p c_s$, with $c_s$ the solute concentration at the entrance of the channel, see Ref.~\cite{Bacchin2022} for details. For salts at a concentration $c_s = 0.1~$mM, $D_s \simeq 10^{-9}$~m$^2$\,s$^{-1}$, $p \simeq 2~$mm, and the salt concentration at the tip is expected to reach $\simeq 1$~M only after few weeks,
so we can state that the accumulation of solutes induced by pervaporation in the experiments described in Sec.~\ref{sec:pervapEKharvesting} does not affect the leaf pervaporation rates (see Ref.~\cite{Bacchin2022} for more informations).

\subsection{Characterisation of the EK conversion element\label{subsec:characEKconversionelement}}

As our aim is to harvest measurable electrical power from pervaporation-driven flows at relatively low rates ($Q < 20~\mu$L\,h$^{-1}$), it is necessary for the EK conversion element to have a  hydraulic  resistance $R_h$ which results in a non-negligible pressure drop $\Delta P = R_h Q$ and mechanical power $\mathcal{P}_h = \Delta P Q$. The hydraulic resistance of a colloid plug in a tube of inner diameter $2R_t$ as shown in Fig.~\ref{fig:EKconversionElement}(a) can be estimated by:
\begin{eqnarray}
R_h = \frac{\eta_w L_p}{\pi R_t^2 \kappa_\mathrm{CK}}, \label{eq:permeability}
\end{eqnarray}
$\eta_w$ being the viscosity of water and $\kappa_\mathrm{CK} \simeq 5.7\times 10^{-15}$~m$^2$  the hydraulic permeability estimated by the Carman-Koseny relation  $\kappa_\mathrm{CK} = a^2 (1-\varphi_d)^3/(45 \varphi_d^2)$, assuming random close-packing, i.e.,  $\varphi_d = 0.64$. This simple estimate leads to a length $L_p$ of a few cm in a tube with $R_t = 0.5~$mm to achieve a pressure drop of the order of $\Delta P \sim 1~$bar for flows driven by pervaporation in the PDMS leaves described previously   (Fig.~\ref{fig:PervaporationLeaves}).

\begin{figure}[ht]
\centering
\includegraphics{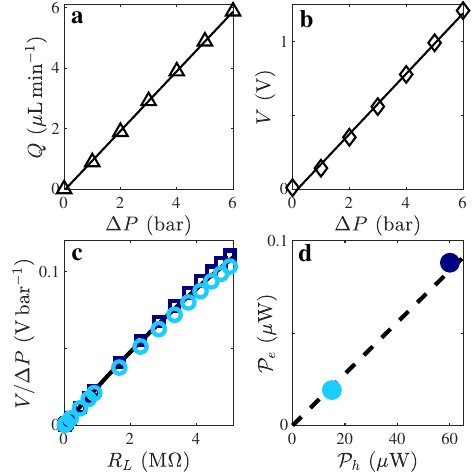}
\caption{Characterisation of the EK conversion element. 
(a)~Flow rate $Q$ across the colloidal plug 
and (b) streaming potential $V$ for $R_L = 10$~M$\Omega$, as a function of the imposed pressure drop $\Delta P$. The continuous lines are linear fits.
(c) Streaming potential $V$ (normalised by  $\Delta P$)
 as a function of the  load resistance $R_L$ ($\circ$ light blue: $\Delta P = 3$~bar,  $\square$ dark blue: $\Delta P = 6$~bar).  The continuous line is the best fit by eqn~\ref{eq:EKlin}, see text. (d) Electrical output power $\mathcal{P}_e$ at $R_L = 5$~M$\Omega$ against the mechanical power $\mathcal{P}_h$. The dotted line shows an efficiency $\epsilon \simeq 0.14\%$.
In these experiments, $L_p \approx 2$~cm, $R_t = 0.5~$mm, and the NaCl concentration is $c_s = 0.1$~mM.} 
\label{fig:EKelementCharacterization}
\end{figure}

Fig.~\ref{fig:EKelementCharacterization}(a) shows the 
measurements of the steady state flow rate $Q$ across a colloidal plug  of length $L_p \simeq 2$~cm in a tube radius $R_t = 0.5$~mm, resulting from imposed pressure drops in  the range $\Delta P=[0$---6$]$~bar. The
 linear behaviour $\Delta P = R_h Q$ with 
$R_h \simeq 1~$bar\,min\,$\mu$L$^{-1}$,
demonstrates that the plug of colloids  does not significantly deform up to $6$~bar and simply acts as a hydraulic resistance, and 
eqn~\ref{eq:permeability} leads to $\kappa_h \simeq 4\times 10^{-15}~\mathrm{m}^2$
close to the Carman-Koseny estimate given above.
Note that we also confirmed in separate experiments without the colloid plug, that $R_h$ largely dominates the  resistance $R_f$ of the filter, tubes, and  the connectors, of the order of  $R_f \approx 1 \times 10^{-3}$~bar\,min\,$\mu$L$^{-1}$.  

Fig.~\ref{fig:EKconversionElement}(b) shows the equivalent electrical circuit of the EK conversion element~\cite{VanDerHeyden2007} connected to a primary conditioning circuit composed of a variable load resistance $R_L$. A pressure-driven flow through the 
 interstices of the colloid plug generates a streaming current $S_\text{str} \Delta P$, $S_\text{str}$ (A\,bar$^{-1}$) being the streaming conductance of the  EK conversion element. A fraction $I$ of this current is collected in the external resistance $R_L$, and the other part corresponds to the electro-migration current through the EK conversion element (of electric resistance $R_C$)  due to the streaming potential $V$. In this classic description, $V$ and $\Delta P$ are related by: 
\begin{eqnarray}
    V = \frac{R_C R_L}{R_C + R_L} S_\text{str} \Delta P\,, \label{eq:EKlin}
\end{eqnarray}
 and the electrical power harvested $\mathcal{P}_e = VI $ is maximal for $R_L = R_C$~\cite{VanDerHeyden2007}.  Fig.~\ref{fig:EKelementCharacterization}(b) reports the measurements of the streaming potential $V$  for a NaCl aqueous solution at concentration $c_s = 0.1$~mM, as a function of the imposed pressure drop $\Delta P$ and for $R_L = 10~$M$\Omega$ (the maximal value of the $R_L$ tested, corresponding to the internal impedance of the multi-meter we used, see Sec.~\ref{subsec:ElectricalFlowMeasurements}).
 The linear behaviour  $V \propto \Delta P$  is expected from eqn~\ref{eq:EKlin}, and our measurements show that the EK conversion element is able to generate a streaming potential of $V \simeq 1$~V  for a pressure drop $\Delta P \simeq~5$~bar. 
 Fig.~\ref{fig:EKelementCharacterization}(c) now shows   $V/\Delta P$ as a function of $R_L$ and  
 for  two imposed pressure drops, $\Delta P=3$ and $6$~bar. The collapse is expected from eqn~\ref{eq:EKlin} and the best fit leads to $S_\text{str} \simeq 25~$nA\,bar$^{-1}$ and $R_C \simeq 25 \pm 5~$M$\Omega$. The estimated value of $R_C$ is associated with a significant  uncertainty since  $R_L < R_C$ in our configuration. 
 The choice of the  salt concentration $c_s = 0.1~$mM comes from experiments similar to those   described in Fig.~\ref{fig:EKelementCharacterization} but for $c_s = 0.01$ and $c_s = 1$~mM, leading to  lower streaming conductances (data not shown). Such a concentration value is also in agreement with the optimal concentration found in experiments on model nanofluidic channels~\cite{VanDerHeyden2007} whose dimensions are a priori close to those of the interstices of our porous medium.

 Fig.~\ref{fig:EKelementCharacterization}(d) finally shows the harvested electric power $\mathcal{P}_e = V^2/R_L$ as a function of the input mechanical power $\mathcal{P}_h = \Delta P Q$ for  $R_L = 5$~M$\Omega$ (corresponding thus to $R_\text{var} = R_\text{app}$, see Sec.~\ref{subsec:ElectricalFlowMeasurements}). These data show that  the maximal power generated at $\Delta P = 6~$bar is about  $\mathcal{P}_e \simeq 0.09~\mu$W and the linear relation  $\mathcal{P}_e = \epsilon \mathcal{P}_h$
indicates an EK conversion efficiency of  $\epsilon \simeq 0.14\%$, while efficiencies up to $3$\% have been reported using  nanofluidic channels~\cite{VanDerHeyden2007} and up to $1.3$\% using a glass microchannel array~\cite{Mansouri2012}. 
Note that in our configuration,  $R_L < R_C$ due the use of a standard  multi-meter with internal impedance which imposes a maximal load resistance of $R_L = 10$~M$\Omega$. We are thus not fully exploiting the efficiency of the EK conversion element, which should be at its maximum for $R_L = R_C \simeq 25$~M$\Omega$. However, we have confirmed the validity of eqn~\ref{eq:EKlin} by carrying out similar experiments but for a colloid plug of length $L_p \simeq~3$~mm and hydraulic resistance $R_h \simeq 0.11~$bar min $\mu$L$^{-1}$, see Fig.~S4. These measurements led this time to an internal resistance $R_C \simeq 4.1$~M$\Omega$ and the harvested electric power  reaches a maximum for $R_L = R_C$, associated to an efficiency of $\epsilon \simeq 0.18\%$. However, such a device would not make it possible to achieve significant pressure drops $\Delta P$ for the proof of concept targeted by our study because the hydraulic resistance is $9$ times lower than that characterised in Fig.~\ref{fig:EKelementCharacterization}(a). 

In the following sections, we do not seek to maximise the conversion efficiency $\epsilon$  by exploring factors such as the nature and size of the colloids, or by maximising the collection of the streaming current by the electrodes (also affecting $\epsilon$ ~\cite{Yaroshchuk2022,Saha2019,Mansouri2012}).
Instead, we focus on the increase of the harvested EK energy $\mathcal{P}_e$  through the increase in input mechanical energy $\mathcal{P}_h = \Delta P Q$ (with  leaves imposing a flow rate $Q$, and the EK conversion element imposing a pressure drop $\Delta P$), and  the limitations of this configuration due to  cavitation.

\subsection{Pervaporation-driven electrokinetic energy harvesting \label{sec:pervapEKharvesting}}

We now discuss the coupling of the PDMS leaves described in Sec.~\ref{sec:secpervap} with the EK conversion element characterized in Sec.~\ref{subsec:characEKconversionelement}. In such experiments, we first imposed a  pressure of $8$~bar on a reservoir of a NaCl solution ($c_s = 0.1~$mM) upstream of the EK conversion element, and we plugged the downstream tube  to $1$ up to $5$ leaves in parallel. Once the leaves are completely filled (no air bubbles trapped in the channels), the  flow rate $Q$ is driven only by water pervaporation because the channels have no outlets, and we released the reservoir to the atmospheric pressure. We then measured the streaming potential $V$ over the  load resistance $R_L$ placed in parallel of the EK conversion element, to estimate the electric power $\mathcal{P}_e = V^2/R_L$,  see Fig.~\ref{fig:PervapPrinciple}.

\begin{figure}[htbp]
\centering
\includegraphics{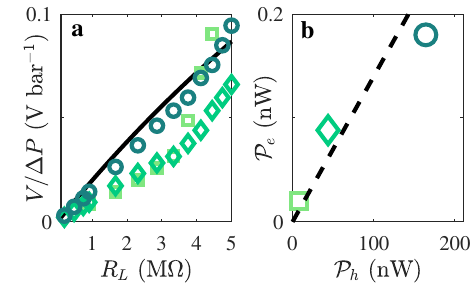}
\caption{Pervaporation-driven EK energy harvesting.
(a)~Streaming potential $V$  
 normalised by the pressure drop $\Delta P$ as a function of the  load resistance $R_L$, the flow is driven by water pervaporation in $1$ ($\square$, light green), $2$ ($\diamond$, green) and $5$ ($\circ$, dark green) PDMS leaves placed in parallel. The continuous line is eqn~\ref{eq:EKlin} with the same parameters as in Fig.~\ref{fig:EKelementCharacterization}(c).
 (b)~Corresponding generated electrical power $\mathcal{P}_e = V^2/R_L$ for $R_L =  5$~M$\Omega$, against the input mechanical power $\mathcal{P}_h$. The dotted line is the linear relation with $\epsilon = 0.14\%$ also found in Fig.~\ref{fig:EKelementCharacterization}(d). 
In these experiments, $L_p \approx 2$~cm, $R_t = 0.5~$mm, and the NaCl concentration is $c_s = 0.1$~mM. 
}\label{fig:pervapEKDATA}
\end{figure}

Fig.~\ref{fig:pervapEKDATA}(a)  shows the measured  data $V/\Delta P$  vs.\ $R_L$, for flows driven by  water pervaporation in $1$, $2$, and $5$ PDMS leaves in parallel. The pressure drop $\Delta P$ across the EK conversion element is estimated using its hydraulic resistance $R_h$ (Fig.~\ref{fig:EKelementCharacterization}(a)) and the measured flow rates $Q$ (Fig.~\ref{fig:PervaporationLeaves}(b)), and varied from $\Delta P = R_h Q \simeq 0.07$~bar for one leaf, to $\Delta P  \simeq 0.32$~bar for $5$ leaves. 
All the data  $V/\Delta P$ against $R_L$ almost collapse on a single curve (despite some dispersion due to the low currents) following again eqn~\ref{eq:EKlin} with the same parameters to those plotted  in Fig.~\ref{fig:EKelementCharacterization}(c) when characterizing the EK conversion element by imposing mechanically a  pressure drop. This shows that the same mechanisms are at play when the flow is imposed mechanically or driven passively by pervaporation.

Fig.~\ref{fig:pervapEKDATA}(b) now reports the output electrical power $\mathcal{P}_e$ generated by the pervaporation-driven flows for $R_L = 5~$M$\Omega$, against the  mechanical power $\mathcal{P}_h =\Delta P  Q $. 
Although the values of $\mathcal{P}_e$ are small, they are measurable and reach $\mathcal{P}_e \simeq 0.18$~nW for the $5$-leaves case. The linear fit shown in Fig.~\ref{fig:pervapEKDATA}(b)
corresponding to the efficiency  $\epsilon = \mathcal{P}_e/\mathcal{P}_h \simeq 0.14\%$ found previously (Fig.~\ref{fig:EKelementCharacterization}(d)),  approximates the experimental data.
It should be noted that such $\epsilon$ measurements have rarely been carried out in the case of evaporation-driven energy harvesting, as this measurement requires not only knowledge of the flow rate of evaporated water, but also the associated pressure drop across the element enabling EK energy conversion. 
As mentioned above, the low electrical power output is closely linked to the low permeability of water through PDMS, but it could also be increased by optimising the collection of the streaming currents by the electrodes, and the nature and size of the colloidal particles used. Nevertheless, the results in Fig.~\ref{fig:pervapEKDATA} provide the first experimental evidence of EK energy conversion from pervaporation-driven flows in PDMS chips. 

\subsection{Pervaporation-induced cavitation in the PDMS leaf}
\begin{figure}[ht]
\centering
\includegraphics{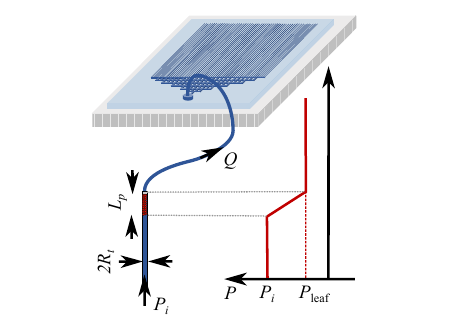}
\caption{The pressure drop $\Delta P$ due to the pervaporation-driven flow rate $Q$ through the colloidal plug leads to a water pressure in the leaf $P_\mathrm{leaf}$ given by eqn~\ref{eq:leafP}. $P_i$ is the  pressure upstream of the plug. 
}\label{fig:pervapPressure}
\end{figure}

Fig.~\ref{fig:pervapEKDATA} shows that the configuration in which the EK conversion element is in series with the flow-driving element offers the possibility pointed out by Yaroshchuk~\cite{Yaroshchuk2022} of  increasing the output power simply   by  increasing, in our case, the pervaporation surface area. However, increasing the output power $\mathcal{P}_e = \epsilon R_h Q^2$ in a passive system by increasing $Q$ can also be limited by  water cavitation caused by a large pressure drop $\Delta P = R_h Q$. This is illustrated in Fig.~\ref{fig:pervapPressure} in the case of a single leaf.
Since the hydraulic resistance $R_h$ of the colloidal plug is much larger than the other hydraulic resistances, including those of the channel network, the  pressure in the leaf  $P_\mathrm{leaf}$, is almost uniform for the flow rates investigated, and given by:
\begin{eqnarray}
P_\mathrm{leaf} \simeq P_i - R_h Q,
\label{eq:leafP}
\end{eqnarray}
where $P_i$ is the  pressure upstream of the EK conversion element. In the data shown in Fig.~\ref{fig:pervapEKDATA}, the water reservoir is opened to the atmospheric pressure ($P_i \simeq 1$~bar) and we found $\Delta P \simeq 0.3$~bar for the $5$-leaves case, leading thus to $P_\text{leaf} \simeq 0.7$~bar,  significantly smaller than  $P_i$ but still positive.

To test the possible cavitation of water for negative leaf pressures, we performed the experiment detailed below and shown  in Fig.~\ref{fig:pervapPressure}. We first imposed a flow of water with a pressure $P_i = 8~$bar through a colloid plug of length $L_p\simeq 6~$cm in a  tube of inner radius $R_t \simeq 0.125~$mm, while measuring simultaneously the resulting flow rate $Q$. This leads to an estimate of the hydrodynamic resistance of the plug, $R_h  \simeq 33$~bar\,min\,$\mu$L$^{-1}$, and 
eqn~\ref{eq:permeability} gives  a    permeability $\kappa_h \simeq 5.9~\times 10^{-15}$~m$^2$, close to the Carman-Koseny estimate $\kappa_\text{CK}$ (see Sec.~\ref{subsec:characEKconversionelement}).  
While still imposing $P_i = 8~$bar, the tube downstream of the plug is connected to a PDMS leaf ($H = 200~\mu$m). Once the air trapped in the dead-end channels have fully permeated through the PDMS layer, the  flow rate $Q$  (now driven only by pervaporation) reaches 
\begin{figure}[ht]
\centering
\includegraphics{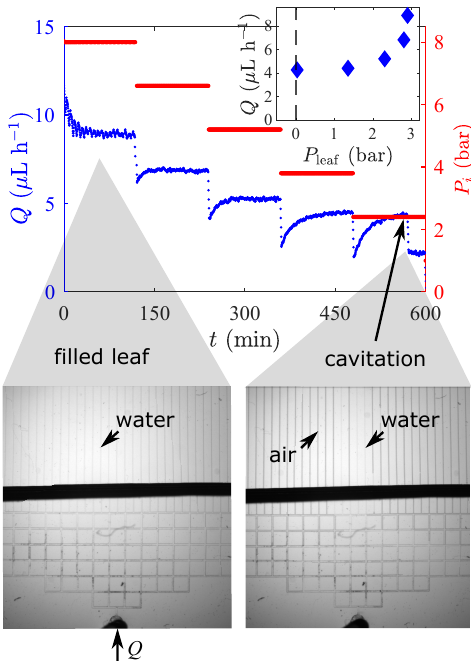}
\caption{Pervaporation-induced cavitation in a PDMS leaf. Flow rates $Q$ vs.\ time $t$ (left axis, blue) measured for different levels of imposed pressure $P_i$ (right axis, red). The inset shows the flow rate value $Q$ at steady state against the pressure in the leaf estimated using eqn~\ref{eq:leafP}. For $t<570$~min, the leaf is filled and water pervaporation induces a flow (bottom left image). At  $t = 570$~min, cavitation occurs in the leaf leading to the formation of bubbles that stop the pervaporation-driven flow (bottom right image), see also Movie S1 in Supplementary information. 
}
\label{fig:cavitation}
\end{figure}
 $Q \simeq 9~\mu$L\,h$^{-1}$  after a transient ($t>30~$min, see Fig.~\ref{fig:cavitation}).
This steady value is then used in eqn~\ref{eq:leafP}  to estimate the leaf pressure $P_\mathrm{leaf} \simeq 3$~bar, see the inset of Fig.~\ref{fig:cavitation} showing the data $Q$ vs.\ $P_\mathrm{leaf}$.

At $t=120~$min, we then imposed lower inlet pressure levels lasting $2$~h, while simultaneously monitoring the channel network using  microscopy at an acquisition frequency of one image every two minutes (Movie S1 in Supplementary information).
For each level of imposed $P_i$, we observed a temporal relaxation of the flow rate to a steady value, which is again used   to estimate  the leaf pressure with eqn~\ref{eq:leafP}. For $P_\mathrm{leaf}>0$,  the steady pervaporation-driven flow rates vary with $P_\mathrm{leaf}$ decreasing from $Q \simeq 9~\mu$L\,h$^{-1}$ at $P_\mathrm{leaf} \simeq 3~$bar to a constant value $Q\simeq 4~\mu$L\,h$^{-1}$ for $P_\mathrm{leaf} \leq 2~$bar.  This variation is actually due to the mechanical deformation of the PDMS layer for high leaf pressures (visible at high magnification), affecting notably the transverse dimensions of both the channel network and the PDMS layer, and thus the pervaporation-driven flow rate $Q$. We confirmed this hypothesis by measuring $Q$ for a PDMS leaf directly connected to a pressure controller to impose $P_\mathrm{leaf}$ (data not shown). These mechanical deformations of the PDMS leaf also explain the  poro-elastic relaxations  
 of the measured flow rate $Q$ at each change in the level of imposed $P_i$, as can be seen in  
Fig.~\ref{fig:cavitation}.

Interestingly, for $P_i = 2.4~$bar ($t>480~$min), the flow rate   first increases up to $Q \simeq 4.3~\mu$L\,h$^{-1}$ corresponding to a leaf pressure $P_\mathrm{leaf} \simeq 0$~bar. Then, cavitation occurs suddenly
 at $t \simeq 570$~min resulting in the formation of bubbles in the PDMS leaf, and stopping the pervaporation-driven flow (see the sudden drop in the measured $Q$ in Fig.~\ref{fig:cavitation}).  
 We repeated this experiment several times with different leaves, but we  always observed  cavitation for   $P_\mathrm{leaf} \simeq 0$~bar. 
 This result is in contradiction with the experiments  of Bruning {\it et al.} who reported a cavitation threshold $P_\text{cap} \simeq -13$~bar, but for a drying water droplet initially trapped in a PDMS mixture during its cross-linking, thus resulting in a defect-free cavity~\cite{Bruning2019}.   
 We thus believe that our observations of systematic cavitation at 
  $P_\text{leaf} \simeq 0$~bar, is due both to the hydrophobic nature of PDMS and to the  defects due to typical microfabrication processes  that can trap  precursors of heterogeneous cavitation~\cite{VincentBook,Loche2024}, 
  despite the initial pressure level at $P_\text{leaf} \simeq 3$~bar for $2$~h (Fig.~\ref{fig:cavitation}) that should eliminate some of the air nanobubbles trapped in the channels.

\section{Conclusion}\label{sec:conclusion}

In the present work, we designed a simple device combining PDMS microfluidic chips and a plug of colloids closely-packed in a tube to provide the first evidence of  EK energy harvesting from pervaporation-driven flows. Although the output electric powers $\mathcal{P}_e$  are low, we have shown that it is possible to achieve higher $\mathcal{P}_e$ simply by increasing the evaporation surface area, and thus the passively-driven flow rate $Q$. 

The versatility and simplicity of the device we have developed offer multiple possibilities for optimisation, in particular by screening for different colloids in the EK conversion element, but also for the ionic strength of the input solution.  
Furthermore, the development of hydrogel-based microfluidic chips
should make it possible to achieve much higher electrical powers, thanks to the greater water permeability of these materials compared to PDMS, and the rules  discussed in Sec.~\ref{sec:secpervap} for designing the channel network of the artificial leaves  should also apply in these cases. 
It should be noted, however, that unlike PDMS, the overall EK energy harvesting performance of hydrogel-based leaves could then be affected by air convection due to their high water permeability.

In our work, we have also demonstrated  that EK energy harvesting  is   limited for high pervaporation-driven flow rates  by water cavitation, which systematically occurs when the pressure drop reaches $\Delta P = R_h Q \simeq 1~$bar corresponding to a water pressure in the PDMS leaves of $P_\text{leaf}  = P_\text{cav} \simeq 0$~bar. 
To significantly reduce the cavitation threshold $P_\text{cav}$ 
in the leaves, it is necessary to use hydrophilic materials, unlike PDMS, and to avoid the presence of surface defects that could trap cavitation precursors~\cite{VincentBook,Loche2024}.  Hydrogel-based chips are possibly good candidates because these materials are intrinsically hydrophilic, and stable flows at significantly negative water pressures (down to $P_\text{cav}= -16.7$~bar)  have been reported despite the inevitable presence of defects due to the microfabrication process~\cite{Liu2023,Wheeler:09}. 
It is important to note that, regardless of the cavitation threshold $P_\text{cav}$ in the element driving the water flow, cavitation a priori always limits the efficiency of the EK energy harvesting in such passive systems as the maximum flow rate cannot exceed $Q = P_\text{cav}/R_h$. This intrinsic limit has never been observed when water evaporation occurs directly from a saturated porous EK energy conversion element, likely due to the very high confinement of water in this case. It is instead capillary phenomena that limit  energy harvesting in such configurations, since a too large pressure drop can lead to a receding of the evaporation surface within the conversion element~\cite{Yaroshchuk2022},  possibly also coupled to the Kelvin effect in the case of nanopores~\cite{Vincent2016}.

Beyond the application studied in the present work, the cavitation threshold measured in the PDMS leaf, together with the variations of the pervaporation-driven flow rate with the applied pressure deforming the channels (see Fig.~\ref{fig:cavitation})
were not reported to our knowledge and could be of interest for the biomimetic studies exploiting pervaporation in  PDMS microfluidic chips to study air propagation in plant leaves after cavitation and embolism formation~\cite{Dollet2021,Keiser2022}.

\section*{Acknowledgements}

This study received financial support from the French government in the framework of the University of Bordeaux’s IdEx “Investments for the Future” program/ GPR PPM. We would like to thank G. Clisson for his technical assistance with the microfluidic experiments, O. Vincent for discussions on heterogeneous cavitation, A.~L. Biance and C. Ybert for discussions of electrokinetic phenomena.


%

\end{document}